\documentclass[aps,prl,twocolumn,superscriptaddress,showpacs]{revtex4-1}
\usepackage{hyperref}

\usepackage{graphicx}  
\usepackage{dcolumn}   
\usepackage{bm}        
\usepackage{amssymb,amsmath}   
\usepackage[usenames,dvipsnames]{xcolor}
\hypersetup{
  colorlinks = true,
  citecolor = {MidnightBlue},
  urlcolor = {MidnightBlue}
}
\usepackage{tikz}
\usepackage{mathtools}
\usepackage[normalem]{ulem}

\usepackage{thermodynamics}

\newcommand{\para}[1]{\paragraph*{#1 ---}}

\newcommand{\with}{\mathrel{;}}

\begin{document}
\title{Marginal and Conditional Second Laws of Thermodynamics}
\author{Gavin E. Crooks}
\email{gec@threeplusone.com} 
\affiliation{Theoretical Institute for Theoretical Science}
\author{Susanne Still}
\email{sstill@hawaii.edu}
\affiliation{University of Hawai$\!$`i at M\=anoa, Department of Information and Computer Sciences and Department of Physics and Astronomy, Honolulu, Hawai$\!$`i 96822, USA}

\begin{abstract}
We consider the entropy production of a strongly coupled bipartite system. 
 The total entropy production can be partitioned into various components, which we use to define local versions of the Second Law that are valid without the usual  idealization of weak coupling.
The key insight is that causal intervention offers a way to identify those parts of the entropy production that result from feedback between the sub-systems. From this the central relations describing the thermodynamics of strongly coupled systems follow in a few lines.
\end{abstract}

\maketitle

Rudolf Clausius' famous statement of  the ``second fundamental theorem in the mechanical theory of heat''  is that ~{``{\sl The entropy of the universe tends to a maximum.}''}~\cite{Clausius1868}
Although this proclamation has withstood the test of time, in practice measuring the entropy of the entire universe is difficult. As an alternative we can apply the Second Law to any system isolated from outside interactions (a Universe unto itself), as, for example, in Planck's statement of the Second Law:~%
``{\sl Every process occurring in nature \ldots the sum of the entropies of all bodies taking part in the process is increased.}''~\cite{Planck1903}
Of course, perfectly isolating any system or collection of systems from outside influence is also difficult. 

Over the last 150 years thermodynamics has progressed by adopting various idealizations which allow us to   isolate and measure that part of the total Universal entropy change that is relevant to the behavior of the system at hand. These idealizations include heat reservoirs, work sources, and measurement devices~\cite{Callen1985,Deffner2013a}. More recently information engines (``Maxwell demons'' \cite{Maxwell1871,Szilard1929}) have been added to the canon to represent idealized computational resources~\cite{Sagawa2010,Horowitz2010,Deffner2013a,Parrondo2015,Boyd2017}. 

In this paper, we demonstrate that we do not need, in principle, to resort to these idealizations. We show how the thermodynamics of strongly coupled systems follow in a straightforward manner from a causal decomposition of the dynamics. This unifying perspective greatly simplifies the treatment, allowing us to assimilate the large recent literature on the thermodynamics of coupled systems in a few short pages. Looking at the problem in the right way then makes it easy to show that conditional and marginalized versions of the Second Law hold locally, even when the system of interest is strongly coupled to other driven, non-equilibrium systems. 

\para{Partitions of entropy} 
Before considering the partitioning of dissipation, let us remember the partitioning of entropy in information theory.  Suppose we have a pair of interacting systems $\X$ and $\Y$, whose states are $x$ and $y$ respectively. The joint entropy of the total system is
\begin{align}
S_{\X,\Y} =- \sum_{x,y} \p(x,y)\ln \p(x,y) \ .
\end{align}
The marginal entropy $S_\X$ of system $\X$ is the entropy of the marginal distribution obtained by summing over the states of the other system,
\begin{align}
S_{\X} &= - \sum_x \p(x) \ln \p(x), \quad p(x) = \sum_y \p(x,y) \ .
\end{align}
The conditional entropy,
\begin{align}
S_{\X\given \Y} 
&= S_{\X,\Y} - S_\Y
 = - \sum_{x,y}   \p(x, y) \ln \p(x\given y) 
\ , 
\end{align}
is the average entropy of system $\X$ given that we know the state of system $\Y$. 
It is also useful to define the pointwise (specific) entropies of individual realizations of the system, whose ensemble averages are the entropies  defined previously. 
\begin{subequations}
\label{pointwise}
\begin{align}
s(x,y) &= - \ln \p(x,y)  = s(x\given y) +s(y) \\
s(x) &= - \ln \p(x)
\label{surprisal}
 \\
s(x\given y) &= - \ln \p(x \given y) 
\end{align}\end{subequations}
The negative log-probability, Eq.~\eqref{surprisal}, is sometimes called ``surprisal" in information theory.

\para{Dissipation}
Let us now consider dynamical trajectories of a bipartite system. We assume that each side of the system is coupled to idealized constant temperature heat reservoirs, with reciprocal temperature $\beta= 1/ k_{\text{B}}T$,  and idealized work sources, the controlled parameters of which we  label $\cx$ and $\cy$. Although the coupling to the heat baths  and work sources are idealized, the two subsystems may be strongly coupled to each other.

A core tenet of non-equilibrium thermodynamics 
is the {\sl detailed fluctuation theorem} which equates the entropy production (or {\sl dissipation}) $\diss$  to the log ratio of the probabilities of forward and time reversed trajectories~\cite{Crooks1998, Jarzynski2000, Seifert2005b}. 
\begin{align}
\diss_{\X,\Y} =
\ln &\frac{\p(\ft{x}, \ft{y} \with \ft{\cx}, \ft{\cy} )}
{\p(\rt{x}, \rt{y} \with \rt{\cx}, \rt{\cy}  )} 
\label{totaldiss}
\end{align}
Dissipation is a consequence of  breaking time-reversal symmetry.
Here, $\ft{x}$ and $\ft{y}$ are trajectories of systems $\X$ and $\Y$ respectively, generated while the systems are driven by the external protocols $\ft{\cx}$ and $\ft{\cy}$ respectively. The trajectory  $\rt{x}$ denotes $\ft{x}$ in reverse, running time backwards. Consequently $\p(\ft{x}, \ft{y} \with \ft{\cx}, \ft{\cy} )$ is the probability of the forward time trajectories, given the forward time dynamics and the forward time protocols, whereas $\p(\rt{x}, \rt{y} \with \rt{\cx}, \rt{\cy} )$ is the probability of the conjugate time reversed trajectories given the time reversed driving.  
We use a semicolon before the controls to emphasize that the protocols are fixed parameters. But for notational simplicity we  typically suppress the explicit dependance of the dynamics on the protocols, writing $\p(\rt{x}, \rt{y})$ for $\p(\rt{x}, \rt{y} \with \rt{\cx}, \rt{\cy} )$, for example.

Suppose we only observe the behavior of one of the subsystems. We can still define marginal trajectory probabilities and a marginal fluctuation theorem, 
\begin{align}
\diss_\X
 = \ln \frac{\p(\ft{x} )}{\p(\rt{x})}  \ ,
\label{marginaldiss}
\end{align}
and, by similar reasoning, the conditional entropy production,
\begin{align}
\label{conddiss}
\diss_{\X|\Y} & =
\ln \frac{\p(\ft{x}\given \ft{y} )}{\p(\rt{x}\given \rt{y} )} 
 = \ln
 \frac{\p(\ft{x}, \ft{y} )}{\p(\rt{x}, \rt{y} )} 
 \frac{\p(\rt{y}  )}{\p(\ft{y} )}
\\ & = \diss_{\X,\Y} - \diss_{\Y} \ .
\notag
\end{align}
Thus we can partition the total dissipation into local components. However, in order to make these definitions of marginal and conditional dissipation concrete we will have to explore their physical significance.

\para{Dynamics}
To make the discussion unambiguous, we  adopt a specific model of the intersystem dynamics. We could opt for classical mechanics~\cite{Deffner2013a}, or coupled Langevin dynamics~\cite{Munakata2014}, or a continuous time Markov process~\cite{Horowitz2014}. But we feel the discussion is most transparent when the dynamics are represented by coupled, discrete time Markov chains.
The dynamics of the joint system are assumed to be Markov, and the dynamics of each subsystem is conditionally Markov given the state of the neighboring subsystems. The marginal dynamics are not Markov when we do not know the hidden dynamics of the other subsystem.

We can use a  causal diagram~\cite{Davis1985,Pearl2009,Ito2013} to illustrate the time label conventions for the trajectories of the system and  control parameters. First one subsystem updates, then the other, and so on. 
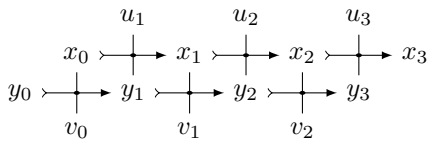
\begin{center}
\begin{tikzpicture}[xscale=0.75,yscale=0.5,baseline=0.75cm]

\node (u1) at (2, 3) {\ensuremath{u_1}};
\node (u2) at (4, 3) {\ensuremath{u_2}};
\node (u3) at (6, 3) {\ensuremath{u_3}};
\node (u01) at (1, 3) {};
\node (u12) at (3, 3) {};
\node (u23) at (5, 3) {};

\node (x0) at (1, 2) {\ensuremath{x_0}};
\node (x1) at (3, 2) {\ensuremath{x_1}};
\node (x2) at (5, 2) {\ensuremath{x_2}};
\node (x3) at (7, 2) {\ensuremath{x_3}};
\node (x01) at (2, 2) {};
\node (x12) at (4, 2) {};
\node (x23) at (6, 2) {};

\node (y0) at (0, 1) {\ensuremath{y_0}};
\node (y1) at (2, 1) {\ensuremath{y_1}};
\node (y2) at (4, 1) {\ensuremath{y_2}};
\node (y3) at (6, 1) {\ensuremath{y_3}};
\node (y01) at (1, 1) {};
\node (y12) at (3, 1) {};
\node (y23) at (5, 1) {};

\node (v0) at (1, 0) {\ensuremath{v_0}};
\node (v1) at (3, 0) {\ensuremath{v_1}};
\node (v2) at (5, 0) {\ensuremath{v_2}};
\node (v01) at (2, 0) {};
\node (v12) at (4, 0) {};
\node (v23) at (6, 0) {};


\draw[>-{latex}]  (x0) edge (x1) ;
\draw[>-{latex}]  (x1) edge (x2) ;
\draw[>-{latex}]  (x2) edge (x3) ;

\draw[>-{latex}]  (y0) edge (y1) ;
\draw[>-{latex}]  (y1) edge (y2) ;
\draw[>-{latex}]  (y2) edge (y3) ;


\draw[] (u1) edge (x01.center);
\draw[] (u2) edge (x12.center);
\draw[] (u3) edge (x23.center);

\draw[] (x0) edge (y01.center);
\draw[] (x1) edge (y12.center);
\draw[] (x2) edge (y23.center);

\draw[] (y1) edge (x01.center);
\draw[] (y2) edge (x12.center);
\draw[] (y3) edge (x23.center);

\draw[] (v0) edge (y01.center);
\draw[] (v1) edge (y12.center);
\draw[] (v2) edge (y23.center);

\fill (x01) circle [radius=1.5pt];
\fill (x12) circle [radius=1.5pt];
\fill (x23) circle [radius=1.5pt];

\fill (y01) circle [radius=1.5pt];
\fill (y12) circle [radius=1.5pt];
\fill (y23) circle [radius=1.5pt];

\end{tikzpicture}\end{center}
Horizontal arrows indicate time evolution, and the other connections indicate causation, where the dynamics of one sub-system are influenced by the external parameters and the current state of the other sub-system.
For the corresponding time reversed trajectory the horizontal arrows flip, but the vertical connections remain unchanged. 
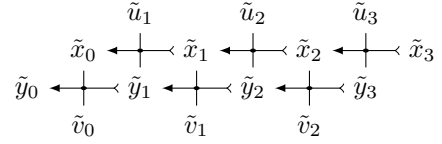
\begin{center}
\begin{tikzpicture}[xscale=0.75,yscale=0.5,baseline=0.75cm]

\node (u1) at (2, 3) {${\rev{u}_1}$};
\node (u2) at (4, 3) {${\rev{u}_2}$};
\node (u3) at (6, 3) {${\rev{u}_3}$};
\node (u01) at (1, 3) {};
\node (u12) at (3, 3) {};
\node (u23) at (5, 3) {};

\node (x0) at (1, 2) {${\rev{x}_0}$};
\node (x1) at (3, 2) {${\rev{x}_1}$};
\node (x2) at (5, 2) {${\rev{x}_2}$};
\node (x3) at (7, 2) {${\rev{x}_3}$};
\node (x01) at (2, 2) {};
\node (x12) at (4, 2) {};
\node (x23) at (6, 2) {};

\node (y0) at (0, 1) {${\rev{y}_0}$};
\node (y1) at (2, 1) {${\rev{y}_1}$};
\node (y2) at (4, 1) {${\rev{y}_2}$};
\node (y3) at (6, 1) {${\rev{y}_3}$};
\node (y01) at (1, 1) {};
\node (y12) at (3, 1) {};
\node (y23) at (5, 1) {};

\node (v0) at (1, 0) {${\rev{v}_0}$};
\node (v1) at (3, 0) {${\rev{v}_1}$};
\node (v2) at (5, 0) {${\rev{v}_2}$};
\node (v01) at (2, 0) {};
\node (v12) at (4, 0) {};
\node (v23) at (6, 0) {};


\draw[>-{latex}]  (x1) edge (x0) ;
\draw[>-{latex}]  (x2) edge (x1) ;
\draw[>-{latex}]  (x3) edge (x2) ;

\draw[>-{latex}]  (y1) edge (y0) ;
\draw[>-{latex}]  (y2) edge (y1) ;
\draw[>-{latex}]  (y3) edge (y2) ;


\draw[] (u1) edge (x01.center);
\draw[] (u2) edge (x12.center);
\draw[] (u3) edge (x23.center);

\draw[] (x0) edge (y01.center);
\draw[] (x1) edge (y12.center);
\draw[] (x2) edge (y23.center);

\draw[] (y1) edge (x01.center);
\draw[] (y2) edge (x12.center);
\draw[] (y3) edge (x23.center);

\draw[] (v0) edge (y01.center);
\draw[] (v1) edge (y12.center);
\draw[] (v2) edge (y23.center);

\fill (x01) circle [radius=1.5pt];
\fill (x12) circle [radius=1.5pt];
\fill (x23) circle [radius=1.5pt];

\fill (y01) circle [radius=1.5pt];
\fill (y12) circle [radius=1.5pt];
\fill (y23) circle [radius=1.5pt];

\end{tikzpicture}\end{center}
Here the tilde $\rti{x}$ labels the time reversed configurations of the time reversed trajectory. 

The probability for the joint trajectory (which is a product of individual transition probabilities) naturally splits into a product of two terms. 
\begin{align}
\p(&\ft{x},\ft{y}  \given  \fti{x}, \fti{y} )   \notag
\\ =  \notag &
\trans{y_0}{y_1}{x_0}
\trans{x_0}{x_1}{y_1}
\trans{y_1}{y_2}{x_1}
\trans{x_1}{x_2}{y_2}
 \\  \notag &
\quad \ldots
\trans{y_{\tau-1}}{y_\tau}{x_{\tau-1}}
\trans{x_{\tau-1}}{x_\tau}{y_{\tau}}
\\ \notag  = &
 \prod_{t=0}^{\tau-1} \trans{y_t}{y_{t+1}}{x_{t}} 
 \times 
\prod_{t=0}^{\tau-1} \trans{x_t}{x_{t+1}}{y_{t+1}}
\\   =  & 
\q(\ft{y} \with  \ft{x}, \fti{y} )  \quad \qquad \times   \q(\ft{x} \with \ft{y}, \fti{x}  )  
\label{causalpartition}
\end{align}
Here, $\p({x_{t+1}}|{x_{t}},{y_{t+1}})$  is the probability of jumping from state $x_{t}$ to $x_{t+1}$  given the current state of the other subsystem  (and given knowledge of the driving protocol).
The expressions $ \q(\ft{y} \with  \ft{x}, \fti{y} )$ and $ \q(\ft{x} \with \ft{y}, \fti{x}  )$ are the trajectory probabilities of one system given a fixed trajectory of the other system. Once again we set off parameters that are fixed (rather than observed and coevolving) with a semicolon. 
This can be represented by the following pair of diagrams:
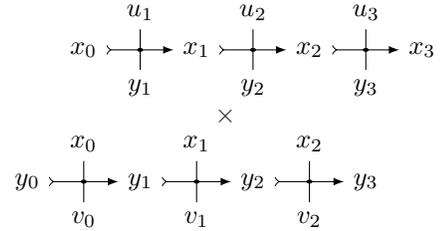
\begin{center}
\usetikzlibrary{arrows}
\begin{tikzpicture}[xscale=0.75,yscale=0.5, baseline=1.0cm]]

\node (u1) at (2, 3) {$u_1$};
\node (u2) at (4, 3) {$u_2$};
\node (u3) at (6, 3) {$u_3$};
\node (u01) at (1, 3) {};
\node (u12) at (3, 3) {};
\node (u23) at (5, 3) {};

\node (x0) at (1, 2) {$x_0$};
\node (x1) at (3, 2) {$x_1$};
\node (x2) at (5, 2) {$x_2$};
\node (x3) at (7, 2) {$x_3$};
\node (x01) at (2, 2) {};
\node (x12) at (4, 2) {};
\node (x23) at (6, 2) {};

\node (y0) at (0, 1) {$\phantom{y_0}$};
\node (y1) at (2, 1) {$ y_1$};
\node (y2) at (4, 1) {$ y_2$};
\node (y3) at (6, 1) {$ y_3$};
\node (y01) at (1, 1) {};
\node (y12) at (3, 1) {};
\node (y23) at (5, 1) {};



\draw[>-{latex}]  (x0) edge (x1) ;
\draw[>-{latex}]  (x1) edge (x2) ;
\draw[>-{latex}]  (x2) edge (x3) ;



\draw[] (u1) edge (x01.center);
\draw[] (u2) edge (x12.center);
\draw[] (u3) edge (x23.center);


\draw[] (y1) edge (x01.center);
\draw[] (y2) edge (x12.center);
\draw[] (y3) edge (x23.center);


\fill (x01) circle [radius=1.5pt];
\fill (x12) circle [radius=1.5pt];
\fill (x23) circle [radius=1.5pt];


\end{tikzpicture}

\\ $\times$ \\
\usetikzlibrary{arrows}
\begin{tikzpicture}[xscale=0.75,yscale=0.5, baseline=0.5cm]


\node (x0) at (1, 2) {$ x_0$};
\node (x1) at (3, 2) {$ x_1$};
\node (x2) at (5, 2) {$ x_2$};
\node (x3) at (7, 2) {$\phantom{x_3}$};
\node (x01) at (2, 2) {};
\node (x12) at (4, 2) {};
\node (x23) at (6, 2) {};

\node (y0) at (0, 1) {$y_0$};
\node (y1) at (2, 1) {$y_1$};
\node (y2) at (4, 1) {$y_2$};
\node (y3) at (6, 1) {$y_3$};
\node (y01) at (1, 1) {};
\node (y12) at (3, 1) {};
\node (y23) at (5, 1) {};

\node (v0) at (1, 0) {$v_0$};
\node (v1) at (3, 0) {$v_1$};
\node (v2) at (5, 0) {$v_2$};
\node (v01) at (2, 0) {};
\node (v12) at (4, 0) {};
\node (v23) at (6, 0) {};



\draw[>-{latex}]  (y0) edge (y1) ;
\draw[>-{latex}]  (y1) edge (y2) ;
\draw[>-{latex}]  (y2) edge (y3) ;



\draw[] (x0) edge (y01.center);
\draw[] (x1) edge (y12.center);
\draw[] (x2) edge (y23.center);


\draw[] (v0) edge (y01.center);
\draw[] (v1) edge (y12.center);
\draw[] (v2) edge (y23.center);


\fill (y01) circle [radius=1.5pt];
\fill (y12) circle [radius=1.5pt];
\fill (y23) circle [radius=1.5pt];

\end{tikzpicture}

\end{center}
These  expressions are not the same as the conditional distributions  $\p(\ft{x} | \ft{y}, y_{0} )$, which describe the original process where both systems coevolve and influence each other. We've chosen to use a different symbol ($q$) for these trajectory probabilities to make this distinction abundantly clear. 
The decomposition of the joint probability in Eq.~\eqref{causalpartition} is refereed to as a causal intervention~\cite{Pearl2009}. This decomposition is central to disentangling the direct and indirect effects of intersystem coupling.

\para{Detailed fluctuation theorem} 
For the complete system the total path-wise entropy production consists of the change in the entropy of the environment (due to the flow of heat from the baths) and a boundary term $\Delta s(x, y)$~\cite{Crooks1998, Jarzynski2000, Seifert2005b},
\begin{align}
\label{detailedft}
\diss_{{\X},{\Y}}
 = \Delta s(x, y)   - \beta  \heat_{\X,\Y}(\ft{x}, \ft{y})  \ .
\end{align}
This boundary term is the difference in pointwise entropy between the initial configurations of the forward and reverse trajectories,\begin{align}
\Delta s(x, y) =  - \ln \p(\rti{x}, \rti{y}   )  +\ln  \p(\fti{x}, \fti{y}  ) \ .
\end{align}
Typically, we either assume that the system is initially in thermodynamic equilibrium for both the forward and reversed processes (as we do for the Jarzynski equality~\cite{Jarzynski1997a}), or we assume that the final ensemble of the forward process is the same as the initial, time reversed probabilities of the reversed process, $\p(\rti{x}, \rti{y} ) =   \p(\ftf{x}, \ftf{y}  )$~\cite{Crooks1999a,Seifert2005b}.  However, in general the initial ensembles need not have any simple relationship: for instance we might be observing a short  segment of a much longer driven process.

We also assume that the energy $E$ of the total system consists of the two subsystem Hamiltonians and an interaction term,
\begin{align}
E_{\X,\Y}(x,y \with \cx, \cy) = E_\X(x\with \cx) + E_\Y(y\with \cy) + E^{\inter}_{\X:\Y}(x,y) \ .
\end{align}
The external baths and control parameters couple to the internal states of each system separately and do not couple directly to the interaction energy, which ensures that the source of energy flowing into the system is unambiguous.

The heat is the flow of energy into the system due to interactions with the bath~\cite{Hunter1993, Jarzynski1997a, Crooks1998, Sekimoto1998,Peliti2008b}.
We can split the total heat into the heat flow for each of the two subsystems, $\heat_{\X,\Y} = \heat_\X + \heat_\Y$,
\begin{subequations}
\label{heat}
\begin{align}
\heat_{\X} =  \sum_{t=0}^{\tau-1}  &  \Bigl[ E_\X(x_{t+1} ,  \cx_{t+1} )  + E_{\X:\Y}^{\inter}(x_{t+1} , y_{t+1})
\notag \\     & 
	  \quad  -   E_\X(x_t , \cx_{t+1} ) - E_{{\X:\Y}}^{\inter}(x_t , y_{t+1})  \Bigr ]
	  \ ,
\\
\heat_{\Y} =  \sum_{t=0}^{\tau-1} & \Bigl[ E_\Y(y_{t+1} ,  \cy_{t} )  + E_{\X:\Y}^{\inter}(x_{t} , y_{t+1})
	\notag \\      & \quad  -  E_\Y(y_t, \cy_{t} ) - E_{\X:\Y}^{\inter}( x_{t}, y_t ) \Bigr ]
	\ .
\end{align}
\end{subequations}

\para{Local detailed fluctuation theorems } 
If the trajectory of system $\Y$ is fixed, then its dynamics act as an idealized work source to system $\X$, and we can write down a standard  fluctuation theorem for  system $X$ alone.
\begin{align}
\ln \frac{\q( \ft{x} \with \ft{y}, \fti{x} )  }{\q( \rt{x} \with \rt{y}, \rti{x} ) } 
\frac{\p(\fti{x}) }{ \p(\rti{x}) } 
=  \Delta s(x) -\beta \heat_\X 
\label{localdiss} 
\end{align}
This is the fluctuation theorem we would obtain were there no feedback from $\X$ to $\Y$. With feedback, this quantity no longer correctly describes the entropy production of sub-system X. Assuming that it does leads to apparent contradictions that would imply that the Second Law has to be modified~\cite{Sagawa2010}. 

What is the quantity that correctly describes the entropy production of sub-system X while coupled to the co-evolving sub-system Y? The answer depends on what information the observer has at hand. If the observer knows the co-evolving state of sub-system Y at all times, then the conditional entropy production, $\Sigma_{X|Y}$, Eq. (\ref{conddiss}) best describes the dissipation encountered by system X alone. In the absence of this knowledge, the marginal $\Sigma_X$, Eq. (\ref{marginaldiss}) describes the entropy production of sub-system~X.


While we have no guarantee that the average of Eq. (\ref{localdiss}) is positive when there is feedback between the systems, we do know that both the marginal and the conditional entropy production obey the Second Law, because they can be written as Kullback-Leibler divergences, which means that they are all non-negative quantities \cite{Gaspard2004b,Jarzynski2006a, Kawai2007a, Gomez-Marin2008c}. 

Therefore, let us now, consider a decomposition of the marginal dissipation: 
\begin{subequations}\begin{align}
&\diss_\X 
=  \ln \frac{\p(\ft{x} )}{\p(\rt{x}  )} 
= \ln \frac{\p(\ft{x}, \ft{y}  )} {\p(\rt{x} , \rt{y}  )} 
	\frac {\p(\rt{y} \given \rt{x}  )}{\p(\ft{y} \given \ft{x}  )}   
\label{a}
\\  
& = \ln  \frac{\p(\ft{x}, \ft{y} \given  \fti{x}, \fti{y}  ) } {\p(\rt{x}, \rt{y} \given  \rti{x}, \rti{y}  ) } 
	 \frac{\p(\fti{x}, \fti{y}  ) } {\p(\rti{x}, \rti{y}  ) } 
  	\frac {\p(\rt{y} \given \rt{x}, \rti{y}  )}{\p(\ft{y} \given \ft{x}, \fti{y}  )} 
 \frac  {\p(\rti{y} \given \rti{x}  ) } 	{\p(\fti{y} \given \fti{x}  ) }  
\label{b}
\\
& = 
\ln
\frac{ \q(\ft{y} \with  \ft{x}, \fti{y} )  } { \q(\rt{y} \with  \rt{x}, \rti{y} )    }
\frac{  \q(\ft{x} \with \ft{y}, \fti{x}  )}{\q(\rt{x} \with \rt{y}, \rti{x}  )}
\frac {\p(\rt{y} \given \rt{x}, \rti{y}  )}{\p(\ft{y} \given \ft{x}, \fti{y}  )}  
\frac{ \p( \fti{x}) } { \p(\rti{x}) }  
\label{c}
\\ 
& = 
\ln   \frac{ \p( \fti{x}) } { \p(\rti{x}) }
+ \ln
\frac{  \q(\ft{x} \with \ft{y}, \fti{x}  )}{\q(\rt{x} \with \rt{y}, \rti{x}  )}
\label{d}
\\ 
\notag & \qquad \qquad \qquad\qquad\qquad\quad
- \ln 
\frac {\p(\ft{y} \given \ft{x}, \fti{y}  )}{\p(\rt{y} \given \rt{x}, \rti{y}  )}
\frac{\q(\rt{y} \with \rt{x}, \rti{y}  )}{  \q(\ft{y} \with \ft{x}, \fti{y}  )}  
\\ 
 & =
\Delta s_\X \qquad -\beta \heat_{\X}   \qquad - \tdiss_\X  
\label{transferdiss}
\end{align}
\end{subequations}
We write down in Eq. \eqref{a} the definition of the marginal dissipation~(see Eq. \eqref{marginaldiss}), and expand it using the definition of conditional probability $p(a,b) = p(a|b) p(b)$. Then we split out the initial state probabilities in Eq. \eqref{b}, and in Eq. \eqref{c} split the probability of the joint trajectory into components without feedback~(see Eq. \eqref{causalpartition}). In Eq. \eqref{d} we gather terms, and in Eq. \eqref{transferdiss} 
realize from this decomposition, that the marginal entropy production of sub-system X is comprised of the contribution without feedback (see Eq. \ref{localdiss}), together with a term that arises as a consequence of the feedback. We call this term $\tdiss_\X$ the {\sl transferred dissipation}. If system $\Y$ did not influence the behavior of system $\X$
then $\q(\ft{y} \with  \ft{x}, \fti{y} )$ would equal $\p(\ft{y} \given \ft{x} , \fti{y} )$ (and similarly for the time reversed components), and the transferred dissipation would be zero. 

With this insight, we can now appreciate the fact that causal intervention, and decomposition of the joint probability of the system as a whole into disconnected parts (Eq. \ref{causalpartition}) is crucial in thinking about feedback systems, because it allows us to decompose the entropy production, revealing the contribution due to feedback.

To summarize, 
the menagerie of local detailed fluctuation theorems~(Eqs. \ref{totaldiss}-\ref{conddiss}) can be expressed solely in terms of differences in local pointwise entropies, Eq. \eqref{pointwise}, the heat flow, Eq. \eqref{heat}, and the transferred dissipation.
\begin{subequations} 
\label{heatdiss}
\begin{alignat}{5}
\diss_{\X\Y} &  = 
 \Delta s_{\X\Y}  &&- \beta  \heat_\X &&- \beta  \heat_\Y
\label{heattotdiss} 
\\ 
\diss_\X & = 
 \Delta s_X &&- \beta \heat_\X &&- \tdiss_\X
\label{heatmargdiss} 
\\ 
\diss_{\X|\Y} & = 
  \Delta s_{\X|\Y}  &&- \beta \heat_\X && + \tdiss_\Y
\label{heatconddiss} 
\end{alignat}
\end{subequations}

\para{Transferred dissipation}
Transferred dissipation can be further decomposed into time-forward and time-reversed components (analogous to the decomposition of the entropy production rate into the Shannon entropy rate and a time reversed entropy rate~\cite{Gaspard2004b}).
\begin{align}
&
\tdiss_\Y
 =  \ln 
 \frac {\p(\ft{x} \given \ft{y}, \fti{x}  )}{\p(\rt{x} \given \rt{y}, \rti{x}  )} 
\frac{\q(\rt{x} \with \rt{y}, \rti{x}  )}{  \q(\ft{x} \with \ft{y}, \fti{x}  )}
 \notag 
\\ 
& =  \ln 
\frac {\p(\ft{x} \given \ft{y}, \fti{x}  )} {\p(\rt{x} \given \rt{y}, \rti{x}  )}
\frac{  \q(\ft{y} \with \ft{x}, \fti{y}  )}{\q(\rt{y} \with \rt{x}, \rti{y}  )}
\frac{\p(\rt{y}, \rt{x} \given \rti{x}, \rti{y})}{\p(\ft{y}, \ft{x} \given \fti{x}, \fti{y})}
\notag 
\\
& = 
 \ln
\frac  {\q(\ft{y} \with \ft{x}, \fti{y}  )}{\p(\ft{y} \given \fti{y}  )} 
- \ln
\frac{\q(\rt{y} \with \rt{x}, \rti{y}  )} {\p(\rt{y} \given \rti{y}  )} 
\notag 
\\
& = \tentropy_{\ft{\Y}}\qquad\qquad - \tentropy_{\rt{\Y}}
\end{align}
After some additional manipulation we recognize that  the first term is the sum of the pointwise {\sl transfer entropies}~\cite{Schreiber2000}. 
\begin{align}
T_{\ft{\Y}}
& = \ln \prod_{t=0}^{\tau-1} \trans{y_t}{y_{t+1}}{x_{t}} - \ln
\prod_{t=0}^{\tau-1} p(y_{t+1}\given y_{0:t} )
 \notag 
\\
& = \sum_{t=0}^{\tau-1} \ln   \frac{p(y_{t+1}\given y_{0:t}, x_{0:t} ) }{ p(y_{t+1}\given y_{0:t})  }
\notag  
\\ & = \sum_{t=0}^{\tau-1}  i(y_{t+1}:  x_{0:t} \given y_{0:t})
\end{align}
Here $i(a:b|c)=\ln p(a,b|c)-\ln p(a|c)p(b|c)$ is the pointwise conditional mutual information, and the slice notation $x_{a:b}$ is shorthand for the sequence $x_a, x_{a+1}, \ldots, x_b$. Thus $T_{\ft{\Y}}$ is the total pointwise transfer entropy from $\X$ to $\Y$ for the forward time trajectory.
Transfer entropy has been investigated as a measure of Granger statistical causality~\cite{Granger1969, Schreiber2000, James2016}.  

Recently, transfer entropy has  been recognized as a component of the thermodynamic dissipation~\cite{Sagawa2012a,Ito2013,Prokopenko2013,Hartich2014,Horowitz2014a}. But transfer entropy can only equal the total transferred dissipation if we construe a process with feedback only in the time-forward dynamics, but no feedback in the time-reversed dynamics.
 Such time-reverse feedback-free reference systems have been studied, e.g.~\cite{Sagawa2010,Horowitz2010}. However, in general we must include the time-reversed component of the transferred dissipation in order to fully appreciate the thermodynamic costs associated with interactions between sub-systems~\cite{Spinney2016}. 

\para{Information Engines}
An interesting  idealized limit to consider is  when the interaction energy is zero $E_{\X:\Y}=0$, but the dynamics are still coupled.  Removing the energetic component to the interaction forces us to carefully consider the role of information flow and computation in thermodynamics, and the relationship between information and entropy~\cite{Sagawa2010,Horowitz2010,Toyabe2010,Sagawa2012a,Sagawa2012b,Hartich2014,Koski2014,Horowitz2015, Parrondo2015,Bartolotta2016,Boyd2017}. 
Although no energy flows between the systems, the transferred dissipation can still be non-zero.
From the point of view of system $\X$, system $\Y$ becomes a purely computational resource (a ``Demon''). This resource  has an irreducible thermodynamic cost which is captured by the transferred dissipation.  Neglecting this cost leads to Maxwell Demon paradoxes where the Second Law appears to be violated.

\para{Local Second Law}

We are now in a position to express the averages of the total, marginal, and conditional dissipation by averages of the quantities we found in our decomposition (Eq. \eqref{heatdiss}). Remember that we can also write them as Kullback-Leibler divergences, and hence each obeys a Second Law like inequality: 
\begin{subequations}
\begin{align}
\bigl\langle{\diss}_{{\X}, {\Y} } \bigr\rangle
& =  \Delta S_{\X,\Y}  - \beta  \average{ \heat_{\X} }- \beta  \average{\heat_{\Y} }
\label{mtdiss} 
\\ 
&=  \sum_{\ft{x},\ft{y}} \p(\ft{x}, \ft{y} ) \ln \frac{\p(\ft{x}, \ft{y} )}{\p(\rt{x}, \rt{y} )} \geq 0
\notag 
\\  
\bigl\langle{\diss}_{{\X}}  \bigr\rangle
  &  = \Delta S_\X - \beta \average{ \heat_{\X} } - \average{\tdiss_{\X} }
 \label{mmdiss} 
\\ 
&=  \sum_{\ft{x}} \p(\ft{x}) \ln \frac{\p(\ft{x} )}{\p(\rt{x} )} \geq 0
\notag 
\\ 
\bigl\langle{\diss}_{{\X} \given {\Y} }  \bigr\rangle
 & = \Delta S_{\X|\Y} - \beta \average{ \heat_{\X} } + \average{ \tdiss_{{\Y}} } 
 \label{mcdiss}
 \\ 
& =  \sum_{\ft{y}}  \p(\ft{y} ) \sum_{\ft{x}}  \p(\ft{x}| \ft{y} ) \ln \frac{\p(\ft{x}\given \ft{y} )}{\p(\rt{x}\given \rt{y} )} \geq 0
 \notag
\end{align}
\end{subequations}

Since the total dissipation is the sum of a conditional and marginal dissipation, it follows that, on average, all of the marginal and conditional dissipations are less than the total dissipation, and we summarize:  
\begin{align}
\bigl\langle{\diss}_{{\X}, {\Y} } \bigr\rangle
 \geq \Bigl\{
 \bigl\langle{\diss}_{{\X}}\bigr\rangle,
 \bigl\langle{\diss}_{{\Y}}\bigr\rangle,
 \bigl\langle{\diss}_{{\X} \given {\Y} }  \bigr\rangle, 
 \bigl\langle{\diss}_{{\Y} \given {\X} }  \bigr\rangle   
\Bigr\} \geq 0
\ .
\end{align}
Thus, strongly coupled systems obey a Local Second Law of Thermodynamics. One has to either consider the dynamics of the system alone, and study the marginal dissipation, or account for the behavior of  other systems directly coupled to the system of interest, and study the conditional dissipation. In either case the  system's average dissipation is non-negative and less than the total dissipation. When studying small parts of the entire Universe, we are allowed to neglect the dissipation occurring elsewhere that is irrelevant to the behavior of the system at hand.

\para{Acknowledgments}
We would like to acknowledge enlightening discussions with David Sivak, Rob Shaw, Tony Bell, Jordan Horowitz, Sosuke Ito, Takahiro Sagawa, and Jim Crutchfield. 
This work was supported by the Foundational Questions Institute (Grant No.\  FQXi-RFP3-1345). 

\bibliography{coupled_systems}

\end{document}